\documentclass[prl,aps,amsmath,amssymb,showpacs,twocolumn,floatfix]{revtex4-1}
\usepackage{graphicx,epsfig}
\usepackage[usenames]{color}
\newcommand{\beq}{\begin{equation}}
\newcommand{\eeq}{\end{equation}}
\newcommand{\bea}{\begin{eqnarray}}
\newcommand{\eea}{\end{eqnarray}}

\newcommand{\ket}[1]{\ensuremath{\left|#1\right\rangle}}
\newcommand{\br}{{\bf r}}

\newcommand{\marton}[1]{#1}
\newcommand{\gergely}[1]{#1}
\newcommand{\balazs}[1]{#1}

\begin{document}
\title{Stabilizing the false vacuum: Mott skyrmions}   
\author{M. Kan\'asz-Nagy$^{1}$, B. D\'ora$^{1,2}$, E. A. Demler$^3$ and G. Zar\'and$^1$}
\affiliation{$^1$BME-MTA Exotic Quantum Phases Research Group, Budapest University of Technology and Economics, Budapest 1521, Hungary}
\affiliation{$^2$Department of Physics, Budapest University of Technology and Economics, Budapest 1521, Hungary}
\affiliation{$^3$Department of Physics, Harvard University, Cambridge, MA 02138, U.S.A}

\begin{abstract}
Topological excitations keep fascinating physicists since many decades. 
While individual  vortices and solitons  emerge and have been observed in many areas of physics, 
their most intriguing  higher dimensional topological relatives,
 skyrmions (smooth, topologically stable textures) and magnetic monopoles -- 
emerging almost necessarily in  any grand unified theory and responsible for charge quantization --
remained mostly elusive.
Here we propose that loading a three-component nematic superfluid such as 
$^{23}$Na into a deep optical lattice and thereby  
creating  an insulating core, one can create topologically stable 
skyrmion textures and investigate their properties in detail.  
We show furthermore that the spectrum of the excitations of the superfluid and their  quantum numbers 
change dramatically in the presence of the skyrmion, and they reflect the presence of a trapped monopole,
as imposed by the skyrmion's topology. 
\end{abstract}

\maketitle

Topological excitations and defects  emerge and play a key role in almost any area of 
physics where spontaneous symmetry breaking occurs. Domain walls, besides
being the necessary ingredients of magnetic recording media, also emerge in cosmology, 
string theory~\cite{cosmic_domain_walls}, \balazs{and} constitute basic  quasiparticles in 1+1 dimensional 
field theories~\cite{rajaraman}. 
Line defects such as vortices belong today  to our basic understanding of superfluidity and 
superconductivity~\cite{tinkham, bloch_review, universe_in_a_He_droplet}, and
dislocations  have a central impact on  the elastic properties of materials 
and on melting and fracture~\cite{dislocations}.

Topological point defects and excitations  such as monopoles~\cite{dirac_monopole,tHooft_monopole,polyakov_monopole} and 
skyrmions  may be even more fascinating than vortices and domain walls. Skyrmions,  originally proposed to describe 
hadronic particles~\cite{skyrme1,skyrme2}, emerge as  smooth, localized, and topologically protected   textures 
of some vector field (order parameter).   Their gauge theoretical counterparts,  the 't Hooft-Polyakov monopoles  
appear  to  be  necessary ingredients of almost  any grand unified theories, and are  needed to 
explain charge quantization.  
While  planar or line-like topological defects are abundant in condensed matter, 
observing topological point defects in the laboratory proved to be very hard, since
only peculiar order parameters support their existence. Moreover, as shown by Derrick~\cite{derrick}, 
skyrmions are generally doomed to shrink to a point or to expand 
to infinity and vanish in a homogeneous system.
It is in fact  only very recently that the existence of monopoles has been convincingly demonstrated 
in spin ice materials such as ${\rm Dy}_2{\rm Ti}_2 {\rm O}_7$~\cite{SpinIce,SpinIceRoderich}, and the spontaneous formation of 
a skyrmion lattice textures has been reported in certain magnetic materials~\cite{skyrmionLattice}. 

\begin{figure}[t]
\includegraphics[width=8cm,clip=true]{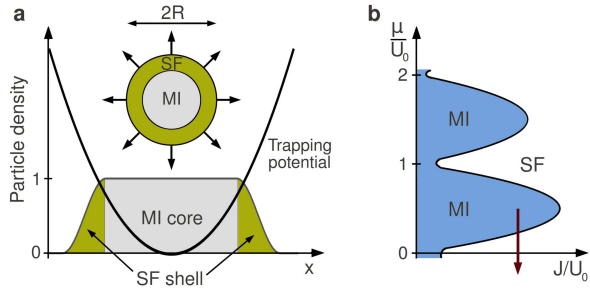}
\caption{\label{fig:trap}
{\bf Schematic structure of the Mott skyrmion. }
(\textbf{a}) A Mott insulator (MI) core is surrounded by a superfluid (SF)
shell with a skyrmion spin structure. Arrows denote the orientation of the nematic order 
parameter, $ \hat{\bf u}(\br)$ in equation \eqref{eq:op}. The MI core stabilizes the skyrmion by 
keeping it from drifting out from the trap.
(\textbf{b}) Schematic finite temperature phase diagram of strongly interacting 
bosons in an optical lattice. The red arrow shows the chemical potential as one 
moves from the center of the trapped skyrmion towards its edge. The MI has no magnetic 
structure at the temperatures considered.
}
\end{figure}

The advent of ultracold atoms opened  new  perspectives to creating and manipulating individual 
skyrmions and monopoles. It has been noticed very early that certain spin $F=1$ multicomponent superfluids
such as $^{23}$Na or $^{87}$Rb support  magnetic  phases~\cite{ho, stoof_nature, demler_zhou, turner_demler, ketterle_nematic},
 which -- combined with an inhomogeneous trap geometry~\cite{oshikawa_herbut}  -- were argued to  give rise 
to  stable skyrmion and  monopole configurations~\cite{stoof_nature, stoof_prl}. 
In a nematic superfluid such as $^{23}$Na, in particular, the superfluid order parameter $ \Psi (\mathbf{r}) = (\Psi_x,\Psi_y,\Psi_z)$ 
takes a remarkably  simple form, 
\beq
\Psi (\mathbf{r}) = \hat{\bf u} (\mathbf{r}) \;\sqrt{\varrho(\mathbf{r}) }  \; e^{i \varphi(\mathbf{r})} \,, 
\label{eq:op}
\eeq
with $\varrho$ denoting the superfluid density, $\varphi$ the superfluid phase, and $\hat{\bf u}$ a real unit vector.
From equation \eqref{eq:op} we can identify  the topological structure  of the nematic order parameter space 
as  $\left(S^2 \times {\rm U}(1)\right)/\mathbb{Z}_2$, with the unit sphere $S^2$ corresponding to the orientation 
of the vector $\hat{\bf u}$ and the ${\rm U}(1)$ symmetry associated with  the phase degree of freedom, $\varphi$.
The curious $\mathbb{Z}_2$ factorization is a consequence of the  fact 
that phase changes  $\varphi \to \varphi+\pi$  are equivalent to flipping the orientation of the vector, 
$\hat{\bf u}\to -\hat{\bf u}$.
 Due to this  structure, topologically  nontrivial and thus topologically stable hedgehog-like  field configurations  
  $\hat{\bf u}(\mathbf{r})$ exist (see Fig.~\ref{fig:trap}) and can give rise to skyrmion and  monopole structures in two and three dimensions, respectively. 
 
Unfortunately however, topological excitations trapped in usual cold atomic 
setups turned out to be  much more  unstable than initially thought;  
monopoles and skyrmions generically slip out  of the trap~\cite{stoof_prl} or 
they simply gradually unwind and disappear within a  short time~\cite{shin}.  
Therefore --  so far -- only unstable skyrmion configurations have been imprinted 
and observed experimentally~\cite{shin, leslie}.

Here we propose to stabilize the skyrmion states \emph{geometrically} 
by generating   a non-superfluid core  at the center of  a trapped nematic superfluid.  We suggest to 
 achieve this by  placing the nematic  superfluid into a deep optical lattice, and 
 thus driving the atoms at the center into a Mott insulating  state  (see Fig.~\ref{fig:trap}).  
In this way a \emph{closed} two-dimensional superfluid shell is created, which -- unlike open shell configurations -- 
supports  topologically stable skyrmions, anchored by the Mott insulating core. 
We compute the free energy of this strongly interacting 'Mott skyrmion' system
 numerically, and demonstrate that the skyrmion texture is indeed stable. 

In the skyrmion configuration the superfluid order parameter acquires a non-trivial, topologically protected 
texture, generated virtually by a \emph{monopole} at the center of the trap.
As predicted by Jackiw and Rebbi and Hasenfratz and 't Hooft~\cite{jackiw,HasenfratztHoft},  the presence 
of a monopole can influence the quantum numbers of the excitations around the monopole --
 and turns bosonic excitations to fermions, a phenomenon termed 'spin from isospin' mechanism. 
As we show by detailed calculations, somewhat similarly to the 'spin from isospin' mechanism, 
the presence of the monopole changes drastically the excitation spectrum of the superfluid \marton{(see also~\cite{Stoof_excitations})}, 
and removes a pseudospin quantum number, present in the 'skyrmionless' ground state. 
 Given the exceptional stability of the 'Mott skyrmion', the predicted 
 change in the excitation spectrum  should be experimentally accessible.  

We describe  a balanced mixture of interacting spin $F=1$  bosons using the lattice Hamiltonian $H_{\mathrm{kin}}+\sum_{\br }H_{\mathrm{loc},\br}$ 
with the kinetic and local parts defined as
\begin{subequations}
\bea
H_{\mathrm{kin}} &=& -J \sum_{\langle \br,\br^\prime\rangle} b_{\br\,\alpha}^\dagger b_{\br^\prime \alpha},
 \\
H_{\mathrm{loc},\br}&=&  -\mu (\br) \,  n_\br +  \frac{U_0}{2}      
:n_{\br}^2: +        \frac{U_2}{2}        :{\vec F}_{\br}^2: \,.     
\eea
\label{eq:H}
\end{subequations}
Here the  operators $b_{\br \alpha}^\dagger $ create a boson of spin component $\alpha$ 
($\alpha = x,y,z$)
at the lattice site $\br$, and $n_{\br} = \sum_\alpha b_{\br\alpha}^\dagger  b_{\br\alpha}$
and  ${\vec F}_{\br} =  \sum_{\alpha,\beta}  b_{\br\alpha}^\dagger \vec{F}_{\alpha\beta} b_{\br\beta}$
denote their density and magnetic moments, respectively. The ${F}^j$  stand for  the usual angular momentum matrices
in  the $\alpha = x,y,z$ basis, 
${F}^j_{\beta\gamma}=-i \, \epsilon_{j \beta\gamma}$, and  $: \ldots :$ refers to normal ordering.
The hopping $J$ sets the kinetic energy of the bosons, 
while the effect of trapping potential $V(\br) = m \omega_{0}^2 \br^2 /2$  is incorporated in the 
effective  position dependent chemical potential, $\mu(\br)  \equiv \mu - V(\br)$.
The (normal ordered) interaction term   $U_0$ describes the strong repulsion between lattice-confined bosons, 
while the second, much weaker  interaction term $U_2$ accounts for the magnetic interaction between them. 
It  is this second term, $U_2>0$, which for  $^{23}$Na forces the superfluid order parameter 
$\Psi_\alpha \propto \langle b_\alpha \rangle   $ 
to stay within the nematic phase, {${\bf f}_\br \equiv \Psi_\br^\dagger \vec F\Psi_\br / |\Psi_\mathbf{r}|^2 \equiv 0$}
(a condition implying  equation~\eqref{eq:op}), 
once the superfluid density ${\varrho }_\br \equiv |\Psi_\br|^2$ becomes  finite.

In the following we focus our attention onto the regime {$zJ/U_0 \approx 0.2$}
with $z=6$ the number of nearest neighbors. Here, increasing the chemical potential at the center of the trap 
beyond some critical value (or equivalently, making the trap tighter),  the density at the center increases
and finally reaches the first Mott lobe (see Fig.~\ref{fig:trap}b). 
For  $^{23}$Na,  in particular, we estimate {$U_0\approx 250 {\rm\, nK}$ and $U_2\approx 6 {\rm\, nK}$} 
in this regime of interest, \gergely{ and {$zJ\approx 50 {\rm\, nK}$}, as shown in Supplementary~Note~1}. 
We also assume that the temperature is already low enough ($T< z J$) to forming a  
superfluid around the Mott core with typical radius $R$, albeit it is still higher than the magnetic ordering temperature of the Mott insulating core, 
$T> T_C\sim J^2/U_0$,  so that the interplay of magnetic ordering in the Mott core  
and superfluidity can be ignored.
 
\section{Results}
 
{\parindent=0pt 
{\bf Stable skyrmion configuration} }
\vskip0.3cm
First, to verify the stability of the skyrmion, we introduced local order parameter fields by performing a Hubbard-Stratonovich 
transformation, $b_\br\to \Psi_\br \approx \langle   b(\br)\rangle$ (see Methods), and 
traced out the original boson fields  numerically to 
obtain the  free energy functional
\bea
\label{eq:free}
F(\{\Psi_\br \} ) &\approx & 
 - J a^2 \sum_{\br,\br^\prime ,\alpha} 
                                        \overline{\Psi}_{\br \alpha} \Delta_{\br\br^\prime } \Psi_{\br^\prime \alpha}
                                        \\
&+& \sum_\br  F_{\rm loc}\left(\varrho_\br,\,  {{\bf f}_\br^{\,2}, \mu(\br),T}\right). 
\nonumber
\eea
Here $a$ denotes the lattice constant, and $\Delta_{\br\br^\prime }$ stands for the discrete Laplace operator. 
The first term describes the stiffness of the superfluid order parameter, while the second term incorporates the effect of the interaction 
as well as that of the confining potential, and has been determined numerically 
for each lattice point  (see Methods). Its  structure follows from 
the obvious $O(3)$ rotational symmetry and $U(1)$ gauge symmetry of the underlying Hamiltonian, 
equation~\eqref{eq:H}.

\begin{figure}[t]
\includegraphics[width=8cm,clip=true]{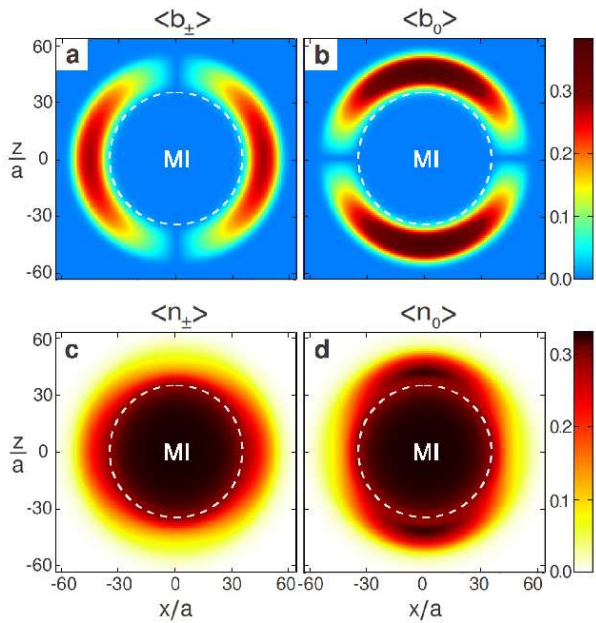}
\caption{ {\bf Inner structure of the skyrmion in the $\mathbf{(x,z)}$ plane.}
(\textbf{a}) In-trap SF densities 
of the $\ket{+1} \, (\ket{-1})$ bosons form a vortex (antivortex)
around the equator, whereas that of the $\ket{0}$ condensate in (\textbf{b}) creates a dark soliton
at the poles.
(\textbf{c},\textbf{d}) show in-trap particle densities. SF order of one of the spin components 
leads to a local increase of the component's particle density
at the expense of those of the other two components, leading to a specific 
density structure characterizing the skyrmion (bottom).
This structure gets significantly more pronounced at lower temperatures.
[Physical parameters of the plot: $T/U_0=0.05,\, U_2/U_0=0.025,\, zJ/U_0=0.18$, 
chemical potential in the middle $\mu_\mathrm{mid}/U_0=0.36$, 
and at the edge $\mu_\mathrm{edge}/U_0=-0.09$.]}
\label{fig:OP}
\end{figure}

To find the minimum of $F(\{\Psi_\br \} )$ we  used the imaginary time equations of motions, 
\beq
-\partial_\tau \Psi_{\br \alpha} = \frac{\delta F}{\delta \overline{\Psi}_{\br \alpha}}, 
\hspace{15 pt}
-\partial_\tau \overline{\Psi}_{\br\alpha} = \frac{\delta F}{\delta \Psi_{\br \alpha}}. 
\label{eq:imag}
\eeq
The  dynamics generated by equation~\eqref{eq:imag} drives the field configuration $\{\Psi_\br \}$ towards the 
minima of the free energy functional. In particular, for appropriate parameters, 
starting from a configuration with a skyrmion texture imprinted, 
 the field $\Psi_\br$  is found to relax to a  configuration 
$$
\Psi_\br \approx    e^{i \varphi} \sqrt{\varrho(r)} \;\hat \br,
$$
with $\hat \br$ denoting the radial unit vector. 
We verified by adding a random component to the initial field configuration 
that this final state is   indeed  a robust local minimum of the free energy, as anticipated.

Fig.~\ref{fig:OP} displays the 
superfluid and total densities across the trap in  the usual hyperfine spin basis, 
$F^z = \pm1$ and  $F^z=0$, where the amplitudes of the various superfluid components 
read  as $\Psi_{\pm}(\br) = e^{i\varphi} \sqrt{\varrho(r)} (\hat{\bf x} \pm i \hat{\bf y})/\sqrt{2}$, and 
  $\Psi_{0}(\br) = e^{i\varphi} \sqrt{\varrho(r)} \hat{\bf z} $.   The superfluid density is clearly  
suppressed at the center of the trap, where the stabilizing Mott insulating core  is formed, and it 
lives on a two dimensional shell around this core.   The components $\Psi_\pm$
form vortices around the equator, while the  $\Psi_0$ component behaves as a dark soliton, 
 localized at the north and south  poles. The total density of the components of the superfluid is also  
 distorted and reflects the structure of the order parameters; the density of 
 $\langle b_0 \rangle \sim \langle b_z \rangle$ is elongated along the $z$-axis, while that of the  other two spin components, 
$\langle b_\pm \rangle \sim \langle b_x \rangle \pm i \langle b_y \rangle$ is squeezed along it (see Fig.~\ref{fig:OP} bottom).

\vskip0.3cm
{\parindent=0pt 
{\bf Creation} }
\vskip0.2cm

A possible way to create the skyrmion is to 
imprint diabatically  a vortex, an antivortex and a 
dark soliton into the three  spin components~\cite{Monopole_creation}, 
and then stabilize the vortex state by turning on a deep optical  optical lattice.
Starting with a superfluid with all atoms in 
the $\ket{-1}$ state, as a first step, a fraction of the atoms could be  
transferred into a vortex state in component $\ket{1}$
using a so-called $\Lambda$ transition.  This is possible by applying a diabatic pulse
of a pair of counter-propagating $\sigma^-,\sigma^+$ Raman beams of
first order Laguerre-Gaussian ($LG^{-1}$) and Gaussian density profiles,
respectively~\cite{leslie}. This  vortex could then be transferred to state $\ket{0}$ by a simple 
RF $\pi$-shift.  As a next step, the creation of an antivortex in component $\ket{1}$ could 
be  achieved by changing the chirality  of the Laguerre-Gaussian beam from $LG^{-1}$  
to $LG^{+1}$. Finally, another laser, perpendicular to the quantization axis, 
would be used to imprint the dark soliton into the remaining
atoms in state $\ket{-1}$. 

An alternative and maybe even simpler way to create the skyrmion could be to imprint three dark solitons 
 in the $x$, $y$, and $z$ directions, respectively, and then mixing them using an RF $\pi/2$-shift.


{\parindent=0pt 
\vskip0.3cm
{\bf Excitation spectrum} }
\vskip0.3cm
To study the excitation 
spectrum of the superfluid shell, we constructed a two-dimensional effective field theory for the superfluid order parameter, $\psi(\br)$ by assuming a  
thin superfluid shell of radius $R$. Neglecting the radial motion of the condensate, we 
can   describe the superfluid by the  Lagrange density 
\beq
{\cal L} = i \,\overline {\psi} \; \partial_t\psi  +  \overline {\psi} \,\bigl (\frac {\Delta_{2}}{2m}+ \tilde\mu\bigr) \psi  - 
\frac { g_0}{2} |\psi|^4 - \frac { g_2}{2} (\overline \psi \vec F \psi)^2 , \label{eq:Lagrangian}
\eeq
generating  the following equations of motion for the order parameter field, 
\beq
i \, \partial_t\psi  =   \,\bigl(-\frac {\Delta_{2}}{2m}- \tilde \mu +  { g_0}|\psi|^2\bigr)\psi  
 +  { g_2}  (\overline \psi \vec F \psi) \cdot \vec  F \psi .
 \label{eq:EOM}
\eeq
Here $\Delta_2$ denotes the two dimensional Laplace operator on the sphere, $\tilde \mu$ the effective 
chemical potential of the superfluid shell, and and $ g_0$ and
$ g_2$ stand for the effective  couplings. All of these parameters depend on the precise width of the 
superfluid shell as well as on the lattice parameters. We estimated them from 
our  lattice computations, \gergely{as explained in more detail in Supplementary~Note~3.}

\begin{figure}[t]
\includegraphics[width=8cm,clip=true]{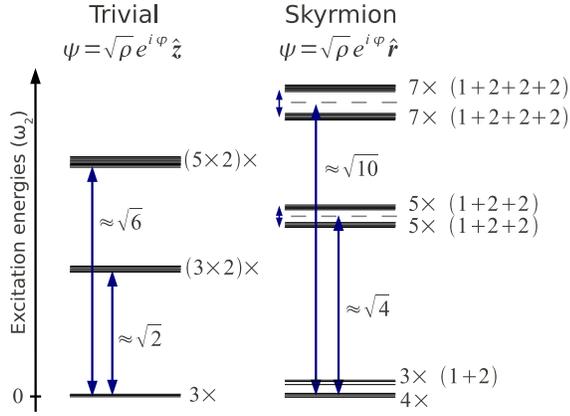}
\caption{\textbf{Excitation spectrum.}  The left (right) panel visualizes the low energy part 
of the excitation spectrum above the trivial (skyrmion) ground states in units of $\omega_2 = 1/(m R \xi_2)$, 
with the magnetic healing length $\xi_2 = 1/\sqrt{m g_2 \rho}$.
Due to the rotational symmetry of the trivial state around $\mathbf{\hat{z}}$, each excited state 
has a 2-fold spin degeneracy in addition to 
the $(2l+1)$-fold orbital degeneracy. In contrast, spin degeneracy splits, 
and only orbital (rotational) degeneracies survive in the skyrmion sector.
The trivial and skyrmion states exhibit different number of zero modes (Goldstone modes) as well.
Apart from the phase degree of freedom, only two zero modes exist in the trivial case, 
since rotations around $\mathbf{\hat{z}}$ in configuration space leave this state invariant.
In the skyrmion state, however, the number of zero modes increases by one, since
rotations around all three spin axes provide a zero mode on top of the phase mode.
\label{fig:spectrum}}
\end{figure}

The excitation spectrum of the condensate  is obtained by linearizing equation~\eqref{eq:EOM} around the ground state 
field configuration, and then solving the resulting coupled differential equations. Equivalently, we can treat the 
field $\psi$ as a quantum field, and obtain the corresponding Bogoliubov spectrum of the condensate \gergely{(see Methods and
Supplementary~Note~4)}. For a uniform, 'skyrmionless' configuration, $\psi\propto \sqrt{\varrho_0}\;\mathbf{\hat z}$, we find that the
density (phase) and spin excitations decouple and the spectrum can be obtained analytically, similarly to the 
\gergely{case of a spatially homogeneous systems}~\cite{spinorreview}. In the limit of large 
 trap radii  compared to the superfluid and magnetic healing lengths, defined through $\xi_{0}^{-2} \equiv {m \,  g_{0} \,\varrho_0} $ 
 and $\xi_{2}^{-2} \equiv {m \,  g_{2} \,\varrho_0} $,  we obtain the spectrum
\bea
\omega_{\rm ph, l} \approx  \frac {1}{mR \xi_0}\sqrt{l(l+1)}\;,
\phantom{nn}
\omega_{{\rm sp},l} \approx    \frac {1}{mR \xi_2}\sqrt{l(l+1)} \;,
\nonumber
\eea  
with $l=0,1,..$ the angular momentum quantum number. Since $ g_2 \ll  g_0$ and thus 
$\xi_2\gg \xi_0$, 
spin excitations dominate the low energy  excitation spectrum of the condensate. 
For a spherical trap every excited state in the spin sector 
has a $(2l+1)\times 2$-fold orbital degeneracy.
 The $(2l+1)$-fold degeneracy is due to spherical symmetry and is accidental in the sense that it is lifted once the trap is distorted, and 
spherical symmetry  is broken. The other, two-fold degeneracy is, however, a consequence of the 
residual $O(2)$ symmetry of the vector part of the order parameter, and can be interpreted as 
a spin degeneracy of the Bogoliubov quasiparticles in the spin sector. Notice that this spin degeneracy is absent in the 
phase sector, where excitations have only a $(2l+1)$-fold angular momentum degeneracy for a 
spherical trap. In addition to the finite energy excitations, three zero-energy excitations (Goldstone modes) 
are found with quantum numbers $l=0$, corresponding to global phase  and 
spin rotations, respectively (see Fig.~\ref{fig:spectrum}).

\marton{Excitations in the skyrmion sector are more complicated, since spin and density fluctuations couple to each other 
due to the spatial winding of the skyrmion texture.} Furthermore, the spherical symmetry is found to be spontaneously broken, and 
the condensate slightly extends into a randomly selected  direction \gergely{(see Supplementary~Note~4)}.
Here we  find four skyrmionic zero-energy Goldstone modes  
and three more excitations of almost zero energy, associated with the spontaneous symmetry breaking.
The rest of the excitations have energies $\omega \sim 1/(m\, R \,\xi_{2})$, and they  come in groups of  $(2l+1)$ almost degenerate 
excitations,  split into 
$l$ states of degeneracy 2 and a non-degenerate state, as induced  by the spontaneous cylindrical distortion of the superfluid. 
The skyrmion's excitation 
spectrum  is sketched in Fig.~\ref{fig:spectrum}. Notice that the structure of the excitation 
spectrum is completely different from that of the 'skyrmionless' sector,  
the spin degeneracy of the  spin  excitations  completely  disappears. 

\marton{The above mentioned coupling between density and spin fluctuations 
leads to a clear fingerprint of the skyrmion state in lattice modulation 
experiments. Modulation of the atom tunneling\gergely{, $J$,} excites oscillations in the amplitude of the superfluid order parameter,
which are coupled to the low energy spin excitations of the skyrmion.
Thus, \gergely{even though these modulations do not couple directly to spin degrees of freedom, 
the topological winding of the skyrmion texture gives 
rise to an indirect coupling to spin excitations, leading in effect to 'spin-orbit coupling'.} 
Specifically we consider modulation of the tunneling along one axis 
by varying the intensity of the optical lattice. This corresponds to perturbation in the $l=0$ and $l=2$ angular
momentum channels, and excites the lower $l=2$ branch of excitations of the skyrmion texture 
\gergely{in Fig.~\ref{fig:spectrum} (see Supplementary~Note~5).}
In contrast to the skyrmion case, spin and density fluctuations are completely decoupled in the trivial sector,
and modulation of the tunneling cannot excite any of the low energy spin modes of this state.
Similarly, modulation of the trapping potential along one direction leads to excitations
of the low energy spectrum of the skyrmion state, but \gergely{no excitation of} the trivial configuration.
We thus find, that the trivial and skyrmion configurations are clearly distinguishable
through analyzing the low energy spectrum of modulation experiments.
}

For typical parameters we find that the energies of the superfluid excitation {are in the order of $10\;{\rm Hz}$.}
It would be essentially impossible to study these excitations in previous  unstable 
skyrmion configurations, however, the extreme stability of our 'Mott skyrmion', should now allow to 
access them and to study their dynamics.

\begin{figure}[b]
\includegraphics[width=8cm,clip=true]{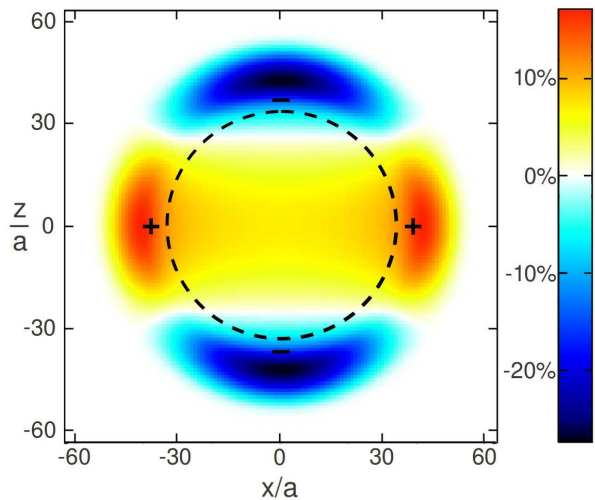}
\caption{ \textbf{Difference of in-trap absorption 
images of the components $\ket{\pm 1}$ and $\ket{0}$, 
taken along the $y$ axis.} Due to the non-trivial SF configurations
of the skyrmion, $\ket{\pm 1}$ ($\ket{0}$) bosons have higher
densities along the equator (poles), see also Fig.~\ref{fig:OP}. 
[Densities are shown as percentages of the largest value of 
the absorption image of component $\ket{0}$. 
Physical parameters: identical to the ones in Fig.~\ref{fig:OP}.] }
\label{fig:diff}
\end{figure}

\begin{figure}[t]
\includegraphics[width=8cm,clip=true]{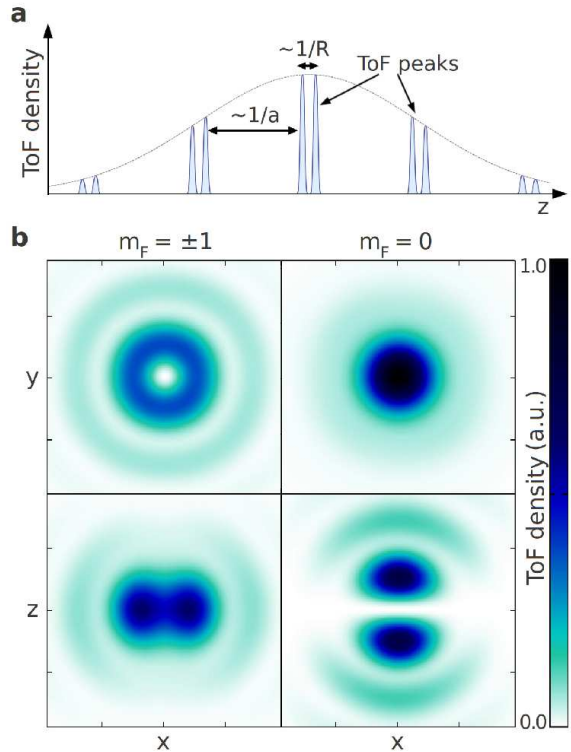}
\caption{\textbf{Time of flight characteristics of the skyrmion.} 
(\textbf{a}) Schematic picture of time of flight (ToF) peaks located at the reciprocal 
lattice vectors in momentum space, with an additional fine structure
($\sim 1/R$) reflecting the spatial SF correlations of the skyrmion
of spatial extent $R$. 
(\textbf{b}) Structure of the doughnut (double peak)
shaped ToF peaks of component $\ket{\pm 1}$ 
($\ket{0}$), on the left (right), 
taken along the $z$ ($y$) axes, top (bottom).
[Color code and axes: arbitrary but identical units.
Physical parameters: identical to those in Fig.~\ref{fig:OP}.] 
\label{fig:ToF}}
\end{figure}

\vskip0.3cm
{\parindent=0pt 
{\bf Detection} }
\vskip0.2cm

The skyrmion texture can be detected in many ways.
Although the change in the density of the components is moderate, 
their in trap density \emph{difference} is rather pronounced and is clearly detectable 
{through absorption imaging} (see Fig.~\ref{fig:diff}).

The skyrmion texture can also be easily  detected through time-of-flight (ToF) measurements, 
imaging the momentum distribution of the trapped atoms~\cite{greiner}.
The ToF image of the atoms with spin component $\alpha$ at time $t$ 
is approximately proportional to~\cite{trivedi}
\beq
n_\alpha^\mathrm{ToF}\propto 
C_{\alpha}\left(\mathbf{k}=\frac{m\mathbf{r}}{ t}\right),
\nonumber
\eeq
with $C_\alpha (\mathbf{k})$ denoting the Fourier transform of the correlation function $\langle b^\dagger_{\br\alpha}  b_{\br^\prime\alpha} \rangle$, 
and is approximately given by 
$C_\alpha (\mathbf{k}) \approx \left| \sum_\br  \langle\Psi_{\br \alpha}\rangle e^{i  \mathbf{k} \br} \right|^2 + const$.
As we show in Fig.~\ref{fig:ToF},  the ToF image consists of Bragg peaks ---- fingerprints of the underlying optical lattice.
Each Bragg peak, however, displays a fine structure at a momentum scale $\mathbf{k}\sim 1/R$ , 
characteristic of the skyrmion texture of the superfluid.   The ToF image of the $m=0$ component, e.g., displays a circular structure when imaged 
from the $z$ direction, while a clear double peak structure should be detected under imaging from the $x$ or $y$ directions.

\vskip0.3cm
{\parindent=0pt 
{\bf Acknowledgement} }
\vskip0.2cm

Illuminating discussions with \'A. Nagy are gratefully acknowledged. 
This research has been  supported by the Hungarian Scientific  Research Funds Nos. K101244, K105149, CNK80991 
and by the Bolyai Program of the Hungarian Academy of Sciences. \gergely{E.~A.~D. acknowledges support through
the DOE (FG02-97ER25308), the Harvard-MIT CUA, the ARO-MURI on Atomtronics, and the ARO MURI Quism program.}

\newpage

\section{Methods}
\textbf{Free energy.}
The partition function is given by 
$Z=\int \mathcal{D}[b^\dagger,b] e^{-S[b^\dagger,b]}$, where the action reads as
\beq
S[b^\dagger,b] = \int d\tau \sum_{\br \alpha} b^\dagger_{\br\alpha} \partial_\tau b_{\br\alpha} 
+ H_\mathrm{kin} + \sum_\br H_{\mathrm{loc},\br},
\nonumber
\eeq
with $H_{\mathrm{kin}}$ and $H_{\mathrm{loc},\br}$ defined in equation \eqref{eq:H}. 
The hopping term is decoupled by a Hubbard-Stratonovich transformation~\cite{fisher_fisher} after introducing
the superfluid order parameter $\Psi$ as 
\bea
&\int & \mathcal{D}[\Psi^\dagger,\Psi] e^{-\sum_{\br\br'}\Psi_\br^\dagger J z \, I^{-1}_{\br\br'} \Psi_{\br'}}
e^{\sum_\br J z (\Psi_\br^\dagger b_\br + b_\br^\dagger \Psi_\br)}
\nonumber\\
&=& e^{\sum_{\br\br'} b^\dagger_\br J_{\br\br'} b_{\br'}},
\eea
with the hopping term expressed as $J_{\br\br'} = J z \, I_{\br\br'}$, with $z = 6$ the number of nearest neighbors.
Its inverse can be rewritten for small kinetic energies using
$I^{-1}_{\br\br'} \approx \delta_{\br\br'} - \frac{a^2}{z} \Delta_{\br\br'}$.
Within the saddle point approximation, the functional integral in $b$ can be carried 
out exactly, leading to the free energy $F[\Psi^\dagger,\Psi] = -T \log Z[\Psi^\dagger,\Psi]$ in equation \eqref{eq:free} with
\bea
&& F_{\rm loc}\left(\rho_\br,\,  {{\bf f}_\br^{\,2}, \mu(\br),T}\right)=  J z \varrho_\mathbf{r} 
 \nonumber \\
& & \phantom{nnn} - T \log \gergely{\mathrm{Tr}_b} \Biggl\{ e^{- \left(H_{\mathrm{loc},\mathbf{r}} 
 - J z (b^\dagger_\mathbf{r} \Psi_{\mathbf{r}} + \mathrm{h.c.})\right)/T } \Biggr\}.
\eea

\textbf{Numerical minimization of the free energy.}
We use a modified version of the imaginary time minimization algorithm implemented
in~\cite{caradoc_davies}. The imaginary time dynamics of the fields in
equation \eqref{eq:imag} lead to a continuous decrease in the free energy 
\beq
\frac{\partial F}{\partial \tau} 
= \sum_\br \frac{\delta F}{\delta \Psi_\br} \frac{\partial \Psi_\br}{\partial \tau}
+ \frac{\delta F}{\delta \Psi_\br^\dagger} \frac{\partial \Psi_\br^\dagger}{\partial \tau}
= -2 \sum_\br \left| \frac{\delta F}{\delta \Psi_\br} \right|^2 < 0.
\nonumber
\eeq

We separate the kinetic part from the remaining on-site contributions
$F=F_\mathrm{kin} + \sum_\br F_{\mathrm{loc},\br}$. $F_\mathrm{kin}$ is calculated using a fast Fourier transform, 
whereas we use numerical interpolation \gergely{in parameter space} to calculate $F_{\mathrm{loc},\br}$ at each site. 
\gergely{For additional details, see Supplementary~Note~2.}


\textbf{Excitation spectrum.}
We analyze excitations around the trivial ($\psi_{t} = \sqrt{\rho_{t}} \; \mathbf{\hat{z}}$) and
skyrmion ($\psi_{s} = \sqrt{\rho_{s}} \; \mathbf{\hat{r}}$) configurations, assuming uniform ground state densities on the sphere.
Here $\mathbf{\hat{z}}$ and $\mathbf{\hat{r}}$ denote unit vectors  in the $z$ and radial directions, respectively.
We determine the two dimensional superfluid densities by using the saddle point approximation for the two dimensional effective Lagrangian $\delta \mathcal{L}/\delta \psi = 0$, yielding 
 $\rho_{t} = \mu /  g_0$ and $ \rho_{s} = (\mu-1/m R^2)/ g_0$, with  $R$ the radius of the sphere.
In the latter case, the chemical potential becomes renormalized by $-1/m R^2$ due to the curvature of the ground state, leading to the 
depletion of the superfluid. 

In the trivial state $\psi_t$  the fluctuations parallel ($\delta\psi_{\parallel}$) and perpendicular 
($\delta\psi_{\perp}$) to the ground state decouple, and their equations of motion 
take on the simple form
\begin{subequations}
\bea
i  \partial_t \delta \psi_{t\parallel} &=& -\frac{\Delta_2}{2 m} \delta\psi_{t\parallel} 
 +   g_0 \rho_{t} \, ( \delta\psi_{t \parallel} + \delta\overline{\psi}_{t \parallel} ), \\
i  \partial_t \delta \psi_{t\perp} &=& -\frac{\Delta_2}{2 m} \delta\psi_{t\perp} 
 +   g_2 \rho_{t} \, ( \delta\psi_{t \perp} - \delta\overline{\psi}_{t \perp} ).
\eea
\label{eq:EOM_methods}
\end{subequations}
The interaction term $g_0$ sets the velocity of 
the fluctuations of the superfluid phase and density,
whereas the spin interaction  $g_2$  determines the velocity of the perpendicular fluctuations.
In the skyrmion state, $\psi_{s} $,  however, the kinetic part in equation \eqref{eq:EOM_methods} 
acquires geometrical terms 
due to the curvature of the ground state configuration. The Laplace operator is 
replaced by
\beq
\Delta_2 \; \rightarrow \; \mathbf{D}^2 + \frac{2}{R^2}, 
\eeq
with  $1/m R^2$ a curvature term shifting the kinetic energy of fluctuations, and  
the covariant derivative ${\mathbf{D}} = {\nabla} + i {\mathbf{A}}$, 
defined using the non-Abelian vector-potential ${\mathbf{A}}$,
mixing the components $\delta\psi_{\perp}\leftrightarrow \delta\psi_{\parallel}$.
These geometrical terms shift the spectrum and 
lead to the splitting of the excitation energies.
\gergely{Further details of this calculation are presented in 
Supplementary~Note~4.}

\newpage
\newpage

\section{Supplementary Information}

\renewcommand{\theequation}{S\arabic{equation}}
\setcounter{equation}{0}  
\renewcommand{\thefigure}{S\arabic{figure}}

\subsection{\gergely{SUPPLEMENTARY NOTE 1}}

\gergely{In this section we provide estimates of the parameters of the lattice Hamiltonian
in equation~(2).} The interaction of spin-1 bosons is determined by their s-wave scattering lengths
$a_0$ and $a_2$, with $a_F$ denoting the scattering length in the total hyperfine spin $F$ channel.
In case of antiferromagnetic interactions, such as in $^{23}\mathrm{Na}$ considered in this paper,
$a_2>a_0>0$ and the ground state is a nematic superfluid~\cite{ho}.
In a deep optical lattice, the on-site interaction
is given by $U_0 = \frac{2\pi^2}{3} \frac{a_0+2 a_2}{\lambda} \left(V_0/E_R\right)^{3/4} E_R$,
with the $V_0$ depth of the optical lattice, and the recoil energy 
of the optical lattice of wavelength $\lambda$ given by 
$E_R = \frac{h^2}{2 m \lambda^2}$~\cite{demler_zhou}.
The hopping, measured in units of $U_0$, is given by
\beq
\gergely{\frac{J}{U_0} = \frac{3}{\pi^{5/2}} \frac{\lambda}{a_0 + 2 a_2} e^{-2 \sqrt{V_0/E_R}}},
\eeq
and can be easily controlled in an experiment by modifying 
the lattice depth to reach the superfluid-Mott insulator \gergely{transition~\cite{demler_zhou,ketterle_scattering_lengths1}}.
The magnitude of the on-site spin interactions is related to $U_0$
by the scattering \gergely{lengths~\cite{demler_zhou, ketterle_scattering_lengths2}},
\beq 
\gergely{\frac{U_2 }{ U_0} = \frac{a_2 - a_0}{a_0 + 2 \, a_2}. }
\eeq
As a result, the interaction term $U_2$ is suppressed, 
and is approximately $U_2 \approx 0.03 \, U_0$ in case of 
$^{23}\mathrm{Na}$~\cite{ketterle_scattering_lengths1,ketterle_scattering_lengths2}.

The parameters used throughout this paper are set using 
the scattering length of $^{23}\mathrm{Na}$, $a_0 = 2.75 \, \mathrm{nm}$,
and a wavelength $\lambda = 594 \, \mathrm{nm}$ for the optical lattice 
as used in the experiment in Ref.~\onlinecite{ketterle_scattering_lengths1}.
In the {$z J / U_0 \approx 0.2$} regime of the phase diagram the parameters become
{$U_0 \approx 250 \, \mathrm{n K}$, $U_2 \approx 6 \, \mathrm{n K}$ 
and $z J \approx 50 \, \mathrm{n K}$},
with $z=6$ the number of nearest neighbors.
Considering $10^5$ atoms in the trap, the radius of the skyrmion is approximately
$R = 30 \, a \approx 10 \, \mathrm{\mu m}$, with 
the lattice constant $a = \lambda / 2 \approx 0.3 \, \mu \mathrm{m}$.

\subsection{\gergely{SUPPLEMENTARY NOTE 2}}
\begin{figure}[t]
\includegraphics[width=8cm,clip=true]{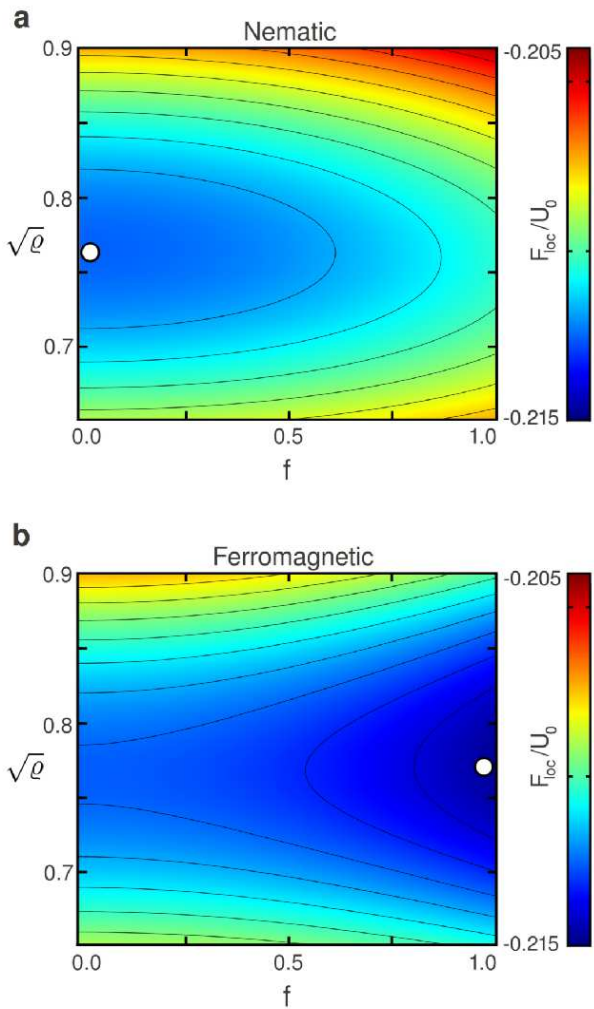}
\caption{\label{fig:free_energies}
\textbf{Local part of the free energy.} $(\textbf{a})$ and $(\textbf{b})$ show $F_{\mathrm{loc}}(\varrho,\mathrm{f})$
in case of nematic ($U_2>0$) and ferromagnetic ($U_2 < 0$) interactions, respectively. 
The dots indicate the minima of the free energy, favoring a non-magnetized (fully magnetized) superfluid 
in the nematic (ferromagnetic) case. 
[Physical parameters of the plot: $T/U_0 = 0.05$, $U_2/U_0 = 0.025$, $zJ/U_0 = 0.40$, $\mu/U_0 = 0.08$.]
}
\end{figure}

\gergely{Here we analyze the structure of the local part of the free energy, 
$F_{\mathrm{loc},\mathbf{r}}$, in equations~(3, 8), and discuss its numerical evaluation.
Due to the $O(3)$ symmetry of the Hamiltonian in equation~(2), $F_{\mathrm{loc},\mathbf{r}}$
can be written as a function of the two rotation-invariant quantities of the $F=1$ spin sector:
the superfluid density ($\varrho_\mathbf{r}$), and
the magnetic moment ($\mathrm{f}_\br = |{\bf f}_\br|$).}
In our numerical simulations, in order to evaluate \gergely{equation~(8)},
we truncate the Hilbert space at $5$ particles per site, and carry out the trace numerically. 
Fig.~\ref{fig:free_energies} shows plots of $F_{\mathrm{loc},\mathbf{r}}$ in
case of nematic ($U_2>0$) and ferromagnetic ($U_2<0$) interactions.
Nematic condensates favor zero magnetization, $\mathrm{f}_\mathbf{r} \equiv 0$,
and thus, the structure of the nematic ground state configuration space is $(S^2 \times U(1))/\mathbb{Z}_2$,
as shown in the main text.
Ferromagnetic condensates, on the other hand, are fully magnetized
$\mathrm{f}_\mathbf{r} \equiv 1$, and thus, their ground state configuration space is $SO(3)$. 
This topological structure, however, holds no topologically protected 'Mott skyrmion' 
configurations~\cite{ho}.

\subsection{\gergely{SUPPLEMENTARY NOTE 3}}

\gergely{In order to get an order of magnitude estimate of the excitation energies 
of the skyrmion, we need to estimate the parameters of the 
two-dimensional effective model in equation~(5).
Therefore, in this part of the Supplementary Information, we relate these parameters 
to those of the lattice Hamiltonian in equation~(2).
Note, that the} accuracy of this estimation does not influence the ratio of the excitation energies 
of the skyrmion and of the trivial sector, shown in Fig.~3.

We approximate the superfluid shell of the skyrmion by a two-dimensional slab 
of thickness $N \approx 10$ lattice sites in the $z$ direction, and we describe it with the action 
\beq
\mathcal{S} = \int dt \sum_{\mathbf{r}\alpha} i \overline{b}_{\mathbf{r}\alpha} \, \partial_{t} b_{\mathbf{r}\alpha}
- \left( H_\mathrm{kin} + \sum_{\mathbf{r}} H_{\mathrm{loc},\mathbf{r}}\right),
\eeq
\gergely{where} the Hamiltonian has been defined in equation~(2).
By assuming a constant, time-independent profile in the $z$-direction for low-energy excitations,
we can approximate the action, using the two-dimensional effective Lagrangian in 
equation~(5),
\beq 
\mathcal{S} \approx N \int dt \int d^2 r \, \mathcal{L}[\overline{\psi},\psi],
\label{eq:Lagrangian_supplementary}
\eeq 
where we have introduced the continuum fields $\psi_\alpha$ through the substitution 
$b_{\mathbf{r}\alpha}/a \to \psi_{\alpha}(\mathbf{r})$, \gergely{using the lattice constant $a$.}
Thus, the parameters of the Lagrangian are given by
$m = 1/(2 J a^2)$, $\tilde{\mu} = \mu + 6 J z$, $g_0 = U_0 \, a^2$ and $g_2 = U_2 \, a^2$.
In the weak coupling limit $z J \gg U_0, U_2$ this Lagrangian density can be used to describe the 
excitation spectrum in the saddle point approximation, as shown in 
the main text. In the strongly interacting limit quantum corrections renormalize the parameters of the Lagrangian density. 
We assume, however, that these effects do not change the order of magnitude of excitation energies significantly,
and therefore we use the bare parameters above to estimate the latter.

The low energy excitations are of the order $E_0 = 1 / ( m R \xi_2 ) = \sqrt{g_2 \rho / m R^2}$ 
both in the skyrmion and in the trivial configuration. Assuming a superfluid 
density $\rho = 0.5 / a^2$, we estimate
\beq
{E_0 \approx \frac{\sqrt{U_2 \, J} }{ R/a} \approx 5 \, \mathrm{Hz}}
\eeq
for the $^{23}\mathrm{Na}$ system \gergely{with the lattice parameters given in Supplementary Note 1}. 
Given the increased stability of the 'Mott skyrmion' considered here,
these frequencies should be in the measurable range.

\subsection{\gergely{SUPPLEMENTARY NOTE 4}}

\gergely{In what follows, we determine and compare the Bogoliubov excitation spectra of the trivial 
and the skyrmion configurations. The excitations were} analyzed using the effective 
Lagrange density in equation~(2), considering a thin superfluid shell of radius $R$.
Assuming spherically symmetric ground state both in the trivial ($\psi_t = \sqrt{\rho_t}  \, \mathbf{\hat{z}}$)
and in the skyrmion configuration ($\psi_s = \sqrt{\rho_s} \, \mathbf{\hat{r}}$), we determined the two-dimensional
superfluid densities using the saddle point equation $\delta \mathcal{L} / \delta \overline{\psi} = 0$. 
This equation yields $\rho_t = \mu / g_0$ and $\rho_s = \left(\mu - 1/(m R^2) \right)/g_0$, respectively.
In the skyrmion case, the chemical potential gets renormalized due to the curvature of the ground state.
This curvature effect leads to the depletion of the superfluid density and affects the excitation spectrum as well.

In the trivial configuration, phase and spin excitations associated with 
the fluctuations parallel ($\delta\psi_{t\parallel}$) and
perpendicular ($\delta\psi_{t\perp}$) to the ground state $\psi_t$, decouple in leading order.
The fluctuation part of the Lagrangian, expanded up to quadratic order,
reads $\delta\mathcal{L} = i\delta\overline{\psi} \, \partial_t\delta\psi - \mathcal{H}$,
with the Hamiltonian density defined as
\beq
\begin{aligned}
\mathcal{H}_t &= \delta\overline{\psi}_t \left( -\frac{\Delta_2}{2m}\, \delta\psi_t \right)
+ g_0 \rho_t \left( |\delta\psi_{t\parallel}|^2 + \frac{\delta\psi_{t \parallel}^2 + \delta\overline{\psi}_{t \parallel}^2}{2} \right) \\
&+ g_2 \rho_t \left( |\delta\psi_{t \perp}|^2 - \frac{\delta\psi_{t \perp}^2 + \delta\overline{\psi}_{t \perp}^2}{2} \right).
\end{aligned}
\label{eq:H_t}
\eeq
The Bogoliubov excitation energies can be obtained by treating the above Hamiltonian quantum mechanically,
or, equivalently, by determining the eigenvalues of the equations of motions of the fields
\bea
i\partial_t \delta\psi_{t\parallel} &=& -\frac{\Delta_2}{2 m} \delta\psi_{t\parallel} 
	+ g_0 \rho_t \left( \delta\psi_{t\parallel} + \delta\overline{\psi}_{t\parallel} \right), 
	\nonumber\\
i\partial_t \delta\psi_{t\perp} &=& -\frac{\Delta_2}{2 m} \delta\psi_{t\perp} 
	+ g_2 \rho_t \left( \delta\psi_{t\perp} - \delta\overline{\psi}_{t\perp} \right).
	\nonumber
\eea
These equations can be easily solved by expanding the fluctuations in terms of spherical harmonics,
yielding the eigenfrequencies
\bea
\omega_{\mathrm{ph},l} &= \sqrt{\left(\dfrac{l (l+1)}{2 m R^2} + g_0 \rho_t\right)^2-(g_0\rho_t)^2}, \nonumber \\
\omega_{\mathrm{sp},l} &= \sqrt{\left(\dfrac{l (l+1)}{2 m R^2} + g_2 \rho_t\right)^2-(g_2\rho_t)^2}, \nonumber
\eea
with the angular momentum quantum number taking values $l=0,1,\dots	$. For a spherical 
trap every excitation in the spin sector has a $(2 l + 1)\times 2$-fold
degeneracy, whereas phase excitations are $(2 l + 1)$-fold degenerate.
Although for a non-spherical trap the $(2 l + 1)$-fold orbital degeneracy is removed, the $2$-fold degeneracy 
of spin modes, a consequence of spontaneous symmetry breaking, remains.
We find three zero-energy excitations (Goldstone modes) with $l=0$
quantum numbers, corresponding to phase fluctuations and rotations of the ground state
around the $x$ and $y$ axes. 
(Rotations around the $z$ axis leave the ground state invariant, 
therefore, they do not give additional zero modes.)
In the limit of large trap radii compared to the superfluid 
($\xi_0 = 1/\sqrt{m\rho g_0}\,$) and magnetic 
healing lengths ($\xi_2 = 1/\sqrt{m\rho g_2}\,$) the excitation energies become
\bea
\omega_{\mathrm{ph},l} &\approx \dfrac{1}{m R \xi_0} \sqrt{l (l+1)}, \nonumber \\
\omega_{\mathrm{sp},l} &\approx \dfrac{1}{m R \xi_2} \sqrt{l (l+1)}. \nonumber
\eea
Since $g_2 \ll g_0$, and thus $\xi_2 \gg \xi_0$, the low energy spectrum is dominated by spin excitations.

In the skyrmion case, the topological structure of the ground state modifies the excitation
spectrum significantly. The kinetic term of the Hamiltonian in equation~\eqref{eq:H_t} acquires a curvature term
$\Delta_2 \rightarrow \Delta_2+ 2/R^2$ due to the non-trivial spatial structure of the ground state.
The equations of motion, therefore, do not decouple and can only be described by the combined equation
\bea
i  \partial_t \delta \psi_{s} &=& -\left(\frac{\Delta_2}{2 m} + \frac{1}{m R^2}  \right)\delta\psi_{s}
  +    g_0 \rho_{s} \, ( \delta\psi_{s \parallel} + \delta\overline{\psi}_{s \parallel} ) \nonumber \\
 &+&   g_2 \rho_{s} \, ( \delta\psi_{s \perp} - \delta\overline{\psi}_{s \perp} ). 
\eea
\gergely{Notice that the action of the seemingly harmless Laplacian is very non-trivial: it mixes 
parallel and perpendicular fluctuations ($\delta\psi_{s\parallel} 
\leftrightarrow \delta\psi_{s\perp}$) due to the skyrmion's geometric structure,
which can also be described by introducing non-Abelian vector potentials, as shown in equation~(10).} 
The excitation energies can be most conveniently found by expanding the fields in the 
(orthonormal) basis of vector spherical harmonic functions~\cite{vector_spherical_harmonics}
\bea
\mathbf{{Y}}_{lm} (\mathbf{r}) &=& \mathbf{\hat{r}} \, Y_{lm}(\mathbf{r}), \nonumber \\
\mathbf{\Psi}_{lm} (\mathbf{r}) &=& r \; \nabla Y_{lm}(\mathbf{r}) / \sqrt{l (l+1)}, \nonumber \\
\mathbf{\Phi}_{lm} (\mathbf{r}) &=& \mathbf{\hat{r}} \times \Psi_{lm} (\mathbf{r}), \nonumber
\eea
defined using the spherical harmonics $Y_{lm}$ of angular momentum quantum numbers $l$ and $m$.
\marton{Due to their vectorial nature, vector spherical functions form a representation of the total angular 
momentum operators ${\vec J} = {\vec L} + {\vec F}$ with quantum numbers $(j,m_J) = (l,m)$,
where the operators ${\vec J}$ account for simultaneous spatial (${\vec L}$) and spin (${\vec F}$) rotations.
}

As can be seen from the formulas above, the vector functions $\mathbf{{Y}}_{lm}$, defined for all $l\ge 0$, always point
in the radial direction; therefore, they span the parallel fluctuations $\delta\psi_{s\parallel}$.
In particular, the function $\mathbf{{Y}}_{00} \propto \psi_s$ corresponds to the skyrmion configuration itself,
and thus, the fluctuation of the corresponding expansion coefficient describes the global phase fluctuations of the skyrmion.
Perpendicular fluctuations, on the other hand, are spanned by the fields $\mathbf{\Psi}_{lm}$ and $\mathbf{\Phi}_{lm}$, 
which are defined for $l = 1, 2, \dots$ angular momenta.

Since the Laplacian leaves the $\mathbf{\Phi}$-sector invariant, 
\beq
-\Delta_2 \mathbf{\Phi}_{lm} = \frac{l(l+1)}{R^2} \mathbf{\Phi}_{lm},
\label{eq:Laplace_phi}
\eeq 
excitations in this sector decouple from the $(\mathbf{{Y}},\mathbf{\Psi})$-fluctuations, and the corresponding
$(2 l + 1)$-fold degenerate excitation energies can be derived analytically,
\beq
\omega_{\mathbf{\Phi},l} = \sqrt{\left( \frac{l(l+1)-2}{2 m R^2} + g_2 \rho_s\right)^2 - (g_2 \rho_s)^2}.
\eeq
In case of large trap radii, $R\gg \xi_0, \xi_2$, these become
\beq
\omega_{\mathbf{\Phi},l} \approx \frac{1}{m R \xi_2} \sqrt{l(l+1)-2}.
\eeq
Specifically, for $l=1$ angular momenta, we find three zero energy modes corresponding to the rotations of the skyrmion around the $x$, $y$ and $z$
axes in parameter space. Therefore, together with the global phase fluctuations in the $\mathbf{{Y}}_{00}$ subspace,
there are four Goldstone modes in the skyrmion sector. We thus find an increased number of Goldstone modes as compared to the trivial sector, 
due to the topological winding of the skyrmion.

The excitation energies of the $(\mathbf{{Y}},\mathbf{\Psi})$-sector are more complicated, 
since the Laplacian is non-diagonal in these fields,
\beq
- \Delta_2 \begin{pmatrix} \mathbf{{Y}}_{lm} \\ \mathbf{\Psi}_{lm} \end{pmatrix} 
= \frac{1}{R^2}
\begin{pmatrix}
l(l+1)+2 & -2 \sqrt{l(l+1)} \\
-2 \sqrt{l (l+1)} & l(l+1) \\
\end{pmatrix}
\begin{pmatrix} \mathbf{{Y}}_{lm} \\ \mathbf{\Psi}_{lm} \end{pmatrix},
\nonumber
\eeq
thereby mixing parallel and perpendicular fluctuations.
The excitation energies are given by the eigenvalues of the 
Bogoliubov-Hamiltonian matrix
\beq
H_{l}^{\mathbf{{Y}}\mathbf{\Psi}} =\frac{1}{2 m R^2} 
\begin{pmatrix} 
\mathbf{\Omega}_l & \mathbf{\Lambda}_m \\
- \mathbf{\Lambda}_m & -\mathbf{\Omega}_l
\end{pmatrix},
\eeq
defined using the matrices
\beq
\mathbf{\Omega}_l = 
\begin{pmatrix}
l (l+1) + \sqrt{2} R / \xi_0 & -2\sqrt{l (l+1)} \\
-2\sqrt{l (l+1)} & l (l+1) - 2 + \sqrt{2} R / \xi_2
\end{pmatrix}
\nonumber
\eeq
and
\beq
\mathbf{\Lambda}_m = (-1)^m \sqrt{2}R \begin{pmatrix} 1/\xi_0 & 0 \\ 0 & -1/\xi_2 \end{pmatrix}.
\nonumber
\eeq
In a spherically symmetric trap there are two branches of excitation energies for all $l=1,2,\dots$ angular momenta,
both being $(2l + 1)$ degenerate. In the $g_2 \ll g_0$ limit the lower of these branches approaches the energies of the corresponding
$\omega_{\mathbf{\Phi}, l} \sim 1/(m R \xi_2)$ spin excitations, whereas the other branch, describing mainly phase excitations,
stays at large energies $\sim 1/(mR\xi_0)$.

An investigation of the $l=1$ excitations reveals a weak instability of the 
spherically symmetric ground state $\psi_s$ towards a slight uniaxial deformation, as
we verified through detailed numerical simulations. This spontaneous 
symmetry breaking does not influence the number of Goldstone modes protected by symmetry,
however, as indicated in Fig.~3, it slightly splits 
the non-zero energy excitations due to the breaking of the $O(3)$ rotational symmetry 
of the ground state to $O(2)$.
In particular, the lower branch of the $l=1$ excitations in the $(\mathbf{{Y}}, \mathbf{\Psi})$-sector 
split in a $3 \rightarrow (2+1)$-manner and their energies become extremely close to 
zero. No such instability has been observed in our three-dimensional lattice simulations, 
though the numerical accuracy of the latter
may have been insufficient to detect this small symmetry breaking.

\subsection{\gergely{SUPPLEMENTARY NOTE 5}}

\gergely{In this section, we analyze the low energy absorption spectrum of the skyrmion in a lattice modulation
experiment, in which atom tunneling along one axis is modulated by periodically varying the depth of the
optical lattice.}
\marton{Specifically, modulations along the $z$ axis correspond to a variation
of the $z$-hopping parameter in equation~(2a). In terms of the two-dimensional effective model
of excitations in equation~(5), this corresponds to a $\partial_z^2$ perturbation operator, as can be seen from the 
discussion below equation~\eqref{eq:Lagrangian_supplementary}. This term has spin $F=0$, and it is a linear
combination of the tensor operators 
\bea
T_{0,0} &=& (\partial_x^2 + \partial_y^2 + \partial_z^2)/\sqrt{3}, \nonumber \\
T_{2,0} &=& (\partial_x^2 + \partial_y^2 - 2 \, \partial_z^2)/\sqrt{6}, \nonumber 
\eea
with angular momentum quantum numbers $(l,m)=(0,0)$ and $(2,0)$, respectively.}

\marton{
The symmetries of our probe operators lead to selection rules for the 
states that can be excited in the spherically symmetric skyrmion configuration, 
$\psi \propto \mathbf{Y}_{00} \propto \hat{\mathbf{r}}$~\cite{footnote_Supplementary}.
Since $T_{0,0}$ and $T_{2,0}$ are derivative operators, they commute with the angular momentum
$L^2 = - \Delta_2$, and they will not mix the subspace $(\mathbf{Y}_{lm},\mathbf{\Psi}_{lm})$ vector spherical harmonics
with $\mathbf{\Phi}_{lm}$ functions, the latter forming an eigenspace of $L^2$ (see equation \eqref{eq:Laplace_phi}).
Further selection rules follow from the rotational symmetries of the perturbation operators under spatial and spin rotations, due to
the Wigner-Eckart theorem~\gergely{\cite{WignerEckart}}. Working in the basis of total angular momentum quantum numbers, it can be easily shown that
the only non-vanishing matrix elements describing excitations of the spherically symmetric skyrmion ground state are 
\bea
\langle \mathbf{Y}_{2,0} | T_{2,0} | \mathbf{Y}_{0,0} \rangle &=& -\sqrt{\frac{2}{15}}, \nonumber \\
\langle \mathbf{\Psi}_{2,0} | T_{2,0} | \mathbf{Y}_{0,0} \rangle &=& \frac{1}{\sqrt{5}}, \nonumber \\
\langle \mathbf{Y}_{0,0} | T_{0,0} | \mathbf{Y}_{0,0} \rangle &=& -\frac{2}{\sqrt{3}}. \nonumber
\eea
Therefore, modulations of the atom tunneling along the $z$ axis can only create $(l,m)=(2,0)$
excitations, and only in the $(\mathbf{Y},\mathbf{\Psi})$ sector. These correspond to a high energy
density excitation, and a small energy spin excitation, the latter being shown in the lower branch of the $l=2$
levels in Fig.~3. 
Such low energy levels cannot be excited in the trivial configuration, whose lattice 
modulation spectrum contains only high energy density fluctuations, to linear order.
Thus, the presence of such a low energy excitation peak in the modulation spectrum 
is an unambiguous fingerprint of the skyrmion texture.
}


\begin{thebibliography}{}
\bibitem{footnote_Supplementary} \marton{Spontaneous symmetry breaking amounts to only $\sim 1\%$ distortion of the skyrmion, 
in the parameter range of $^{23}\mathrm{Na}$ experiments. Thus, its effect on the excitation spectrum will be negligible, 
and we will assume the skyrmion to be spherically symmetric in the following discussion.}
\end{thebibliography}
\end{document}